\documentclass[conference,letterpaper]{IEEEtran}

\usepackage{graphicx}
\usepackage{amsthm}
\usepackage{bm}
\usepackage{times}
\usepackage{mathtools}
\usepackage{amsmath}
\usepackage{amssymb}
\usepackage{bbm}

\usepackage{algorithm}
\usepackage{algorithmicx}
\usepackage{algpseudocode}

\makeatletter
\renewcommand{\fst@algorithm}{%
  \fs@ruled
  \def\@fs@pre{%
    \kern0.08in
    \hrule height.8pt depth0pt
    \kern2pt
  }%
}
\makeatother

\usepackage{tikz}
\usetikzlibrary{positioning,calc}

\usepackage{url}
\usepackage[hidelinks,bookmarks=false]{hyperref}
\usepackage{xcolor}

\algrenewcommand\algorithmicrequire{\textbf{Input:}}
\algrenewcommand\algorithmicensure{\textbf{Output:}}

\date{}

\begin{document}

\title{\huge{Learned Parallel Bit-Flipping Sequential Belief Propagation Decoding of Quantum LDPC Codes}}

\author{
Mohsen~Moradi$^*$,
Taejoon~Kim$^*$,
and
R\'emi~A.~Chou$^\dag$\\
$^*$School of Electrical, Computer and Energy Engineering,
Arizona State University, Tempe, AZ 85287, USA\\ 
Email: \{mmorad11, taejoonkim\}@asu.edu, \\
$^\dag$Department of Computer Science and Engineering, 
The University of Texas at Arlington, Arlington, TX, USA \\
Email: remi.chou@uta.edu
}

\maketitle

\begin{abstract}
Quantum low-density parity-check (QLDPC) codes are promising candidates for low-overhead fault-tolerant quantum computation, but their practical use requires fast, low-complexity, and reliable decoders. Belief propagation (BP) is attractive because of its local message-passing structure, yet standard flooding BP often suffers from convergence failures on QLDPC codes due to short cycles, degeneracy, and symmetric decoding trajectories. Learned sequential BP improves convergence by using a reinforcement-learning policy to choose variable-node update orders, but a single learned trajectory can still be sensitive to unfavorable local Pauli decisions.
We propose a learned parallel bit-flipping sequential BP decoder with a new quaternary score combining syndrome gain, a quantized Pauli log-likelihood penalty, and Q-table lookahead at hypothetical post-flip states.
Our decoder first runs learned sequential BP for a fixed number of iterations. If the syndrome is not satisfied, it constructs quaternary bit-flipping candidates, each corresponding to changing the current Pauli decision of one qubit. The selected candidates initialize independent learned sequential BP continuations from the same decoder state. Since these continuations are independent, they can be executed in parallel, so testing several candidates mainly increases parallel hardware resources rather than sequential decoding latency. Simulations on representative QLDPC codes over the depolarizing channel show that the proposed decoder improves the reliability of learned sequential BP while having a parallel low-latency structure.
\end{abstract}



\section{Introduction}

Quantum error correction is essential for protecting quantum information from decoherence, gate errors, and measurement imperfections. Among quantum error-correcting codes, quantum low-density parity-check (QLDPC) codes have received significant attention because their sparse stabilizer structure enables local syndrome measurements and, for suitable high-rate constructions, can substantially reduce the physical-qubit overhead compared with surface-code architectures \cite{bravyi2024high}. Early QLDPC constructions, most notably hypergraph-product codes \cite{tillich2014quantum}, established explicit positive-rate families. More recent constructions, including balanced-product codes \cite{breuckmann2021balanced}, lifted-product codes \cite{panteleev2022asymptoticallygood}, and quantum Tanner codes \cite{leverrier2022quantum}, have substantially improved the asymptotic rate-distance landscape. In addition, bivariate bicycle (BB) codes provide concrete finite-length QLDPC codes with attractive overhead and threshold behavior, making efficient decoding a pressing practical problem \cite{bravyi2024high}.

Belief propagation (BP) is a natural decoding candidate for QLDPC codes because it is local, iterative, and scalable. However, applying BP directly to QLDPC codes is more challenging than in the classical LDPC setting. The commutativity constraints of stabilizer codes induce many short cycles in the Tanner graph, weakening the independence assumptions behind BP \cite{poulin2008iterative,roffe2020decoding}. A second difficulty is quantum degeneracy: many distinct Pauli error patterns can produce the same syndrome and may correspond to equivalent recovery operations \cite{panteleev2021degenerate}. These effects can cause standard BP to oscillate, stall, or converge to a syndrome-consistent but logically incorrect estimate.

A substantial body of recent work has therefore focused on improving BP-based QLDPC decoders while preserving the locality of message passing. One direction augments BP with post-processing, including BP-OSD \cite{panteleev2021degenerate}, stabilizer inactivation \cite{ducrest2022stabilizer}, and localized statistics decoding \cite{hillmann2025localized}. A second direction modifies the BP dynamics themselves through neural or overcomplete-check BP \cite{miao2025quaternary}, guided decimation \cite{yao2024bpgd}, trapping-set-aware refinements \cite{chytas2024trapping}, layered scheduling \cite{ducrest2023layered}, informed dynamic scheduling \cite{huang2026informed}, and sequential schedules \cite{moradi2026sequential}. These methods show that controlled asymmetry, introduced through post-processing, decimation, or scheduling, can improve BP convergence on QLDPC Tanner graphs.

Scheduling is particularly attractive because it preserves the BP framework while changing how information propagates through the graph. In classical LDPC and polar decoding, the update order can strongly affect convergence, and reinforcement learning has been used to learn effective schedules \cite{zhang2005shuffled,habib2021reldec,moradi2025polar}. Similar ideas have recently been applied to QLDPC decoding. Sequential check-node scheduling (SCNS) and sequential variable-node (VN) scheduling (SVNS) improve BP by immediately reusing the freshest messages within each iteration \cite{moradi2026sequential}. Building on this idea, reinforcement-learning-based sequential BP decoding, we refer to as RL-S, learns a state-dependent VN schedule from a local syndrome-driven state representation \cite{moradi2026learning}. Under depolarizing noise, the quaternary version of RL-S updates the BP messages associated with the two binary components of each Pauli error and uses the learned schedule to guide the decoding trajectory.

Despite these gains, a single learned sequential trajectory can still be vulnerable to an unfavorable local Pauli decision. This is especially relevant under depolarizing noise, where several Pauli labels may have comparable posterior likelihoods for the same qubit. Candidate-generation and list-style approaches provide one way to address this limitation. Recent QLDPC decoders explore multiple BP-based candidates using speculative syndrome flips \cite{wang2025fullyparallel}. 
Related bit-flipping ideas have also been studied for QLDPC codes \cite{chytas2024collective} and for classical coding problems, including FDPC \cite{hosseinzadeh2025fdpc} and PAC \cite{moradi2022pacbf} codes. 
Restart Belief~\cite{valentini2025restart} uses successive error guesses, while BP-GDG~\cite{gong2024gdg} progressively fixes variables. Our score jointly evaluates syndrome reduction, soft Pauli likelihood loss, and learned scheduling values at hypothetical post-flip states. Selected changes initialize one-shot RL-S continuations from the saved BP state without permanent decimation. The underlying unquantized likelihood gap is further developed into a cumulative path metric in our subsequent RL-LS extension~\cite{moradi2026list}, which repeatedly expands and prunes decoding trajectories.


In this paper, we propose a learned parallel bit-flipping sequential BP decoder for QLDPC codes. The decoder first runs RL-S for a fixed number of $T_0$ iterations. If the resulting estimate does not satisfy the syndrome, it evaluates possible single-qubit Pauli changes and selects the top candidates according to a quaternary bit-flipping score. Each selected candidate initializes an independent branch from the same decoder state. In each branch, the decoder applies one Pauli change, refreshes the affected residual checks and BP messages, and runs a short learned sequential BP continuation of $T_{\rm br}$ iterations. Since the branches are independent, they can be implemented as parallel lanes. Thus, the proposed method adds search diversity while keeping the decoding latency close to that of testing only one candidate.

Our main contributions are summarized as follows:
\begin{itemize}
    \item We propose a learned parallel bit-flipping sequential BP decoder for QLDPC codes over the depolarizing channel.
    \item We introduce a quaternary bit-flipping candidate rule that changes the Pauli decision of one qubit and scores candidates using syndrome gain, reliability, and learned scheduling information.
    \item We show that the selected candidates can be tested through independent continuation branches, enabling a parallel low-latency implementation.
    \item We evaluate the proposed decoder on representative QLDPC benchmark codes and compare it with learned sequential BP and other BP-based baselines.
    For the \( [[288,12,18]] \) BB code of \cite{bravyi2024high}, our proposed decoder matches the performance of RL-S with \(T=1000\) using decoding depth \(T_0+T_{\rm br}=200\), and improves over BP and BPGD by more than one order of magnitude at a depolarizing noise of \(p=0.05\).
\end{itemize}


\section{Background}
\label{sec:background}

\subsection{CSS QLDPC Codes and Syndrome Decoding}

Let \(\mathcal{P}_n\) denote the \(n\)-qubit Pauli group. A stabilizer code is specified by an abelian subgroup \(\mathcal{S}\subset \mathcal{P}_n\) that does not contain \(-I^{\otimes n}\), where \(I^{\otimes n}\) is the \(n\)-fold Kronecker power of the identity operator \(I\). The code space is the joint \(+1\)-eigenspace of all stabilizer generators. An \([[n,k,d]]\) stabilizer code encodes \(k\) logical qubits into \(n\) physical qubits and has minimum distance \(d\).

In this paper, we focus on Calderbank--Shor--Steane (CSS) QLDPC codes. A CSS code is described by two sparse binary parity-check matrices
\begin{equation}
    H_X \in \mathbb{F}_2^{m_X\times n}, \qquad
    H_Z \in \mathbb{F}_2^{m_Z\times n},
\end{equation}
satisfying the commutativity condition
\begin{equation}
    H_X H_Z^{\mathsf{T}} = 0
    \quad \text{over } \mathbb{F}_2 .
\end{equation}
The rows of \(H_X\) correspond to \(X\)-type stabilizer checks, and the rows of \(H_Z\) correspond to \(Z\)-type stabilizer checks. The low-density property means that both matrices are sparse, so each qubit participates in only a small number of checks.

Let
\begin{equation}
    \bm{q}=(q_1,\ldots,q_n), \qquad q_i\in\{I,X,Y,Z\},
\end{equation}
denote a Pauli error. We represent each Pauli symbol by two binary components
\begin{equation}
    e^X(q) \triangleq \mathbbm{1}\!\left[q\in\{X,Y\}\right], \qquad
    e^Z(q) \triangleq \mathbbm{1}\!\left[q\in\{Z,Y\}\right].
\end{equation}
Thus, \(e^X(q)\) indicates whether the Pauli symbol contains an \(X\)-component, and \(e^Z(q)\) indicates whether it contains a \(Z\)-component. For a length-\(n\) Pauli error \(\bm{q}\), we write the corresponding binary component vectors as \(e^X\) and \(e^Z\). The measured CSS syndrome is
\begin{equation}
    s^X = H_X e^Z, \qquad
    s^Z = H_Z e^X,
    \label{eq:css_syndromes}
\end{equation}
where all operations are over \(\mathbb{F}_2\). Hence, \(H_X\) detects the \(Z\)-component of the error, while \(H_Z\) detects the \(X\)-component.

A decoder receives \((s^X,s^Z)\) and produces an estimate
\begin{equation}
    \hat{\bm{q}}=(\hat{q}_1,\ldots,\hat{q}_n),
\end{equation}
or equivalently component estimates \((\hat{e}^X,\hat{e}^Z)\). For a tentative estimate, define the residual syndrome mismatches \cite{moradi2026learning}
\begin{equation}
    \delta^X \triangleq s^X \oplus H_X \hat{e}^Z,
    \qquad
    \delta^Z \triangleq s^Z \oplus H_Z \hat{e}^X,
    \label{eq:residuals}
\end{equation}
and the total residual mismatch weight
\begin{equation}
    w \triangleq \|\delta^X\|_1+\|\delta^Z\|_1 .
    \label{eq:mismatch_weight}
\end{equation}
The condition \(w=0\) means that the estimated Pauli error matches the measured syndrome. In quantum decoding, however, syndrome matching alone does not necessarily imply decoding success. The residual error between the true error and the estimate must also be logically trivial, i.e., it must belong to the stabilizer group.

\subsection{Depolarizing Channel and Quaternary BP}

Under the depolarizing channel with physical error probability \(p\), each qubit is independently affected by
\begin{equation}
    I, X, Y, Z
\end{equation}
with probabilities
\begin{equation}
    P(I)=1-p, \qquad P(X)=P(Y)=P(Z)=\frac{p}{3}.
\end{equation}
The two binary components are therefore statistically coupled, since a \(Y\) error contains both an \(X\)- and a \(Z\)-component. A quaternary BP decoder accounts for this coupling by forming beliefs over the four Pauli symbols instead of decoding the two components independently.

Let \(L_i^X\) and \(L_i^Z\) denote the soft information associated with the two binary components at qubit \(i\). From these values and the depolarizing prior, the decoder computes a local log-likelihood score \cite{moradi2026learning}
\begin{equation}
    \Lambda_i(q), \qquad q\in\{I,X,Y,Z\},
\end{equation}
for each Pauli symbol. The hard decision is
\begin{equation}
    \hat{q}_i = \arg\max_{q\in\{I,X,Y,Z\}} \Lambda_i(q).
    \label{eq:hard_decision}
\end{equation}
This quaternary decision rule is important under depolarizing noise because the most likely Pauli symbol generally depends on the joint four-symbol belief, not on two separate binary component decisions.

\subsection{Learned Sequential BP Scheduling}

Our proposed decoder builds on learned sequential BP scheduling of \cite{moradi2026learning}. In conventional flooding BP, all messages of the same type are updated in parallel at each iteration. In SVNS, VNs are updated one at a time, and each new message is immediately available to later updates in the same sweep. This breaks the synchrony of flooding BP and can reduce oscillations on loopy Tanner graphs.

In learned sequential BP, the VN update order is chosen by a reinforcement-learning policy. The learned policy is represented by a Q-table, denoted by \(\mathcal{Q}\), indexed by a local state and a VN index. For each VN \(i\), the local state is determined by the pattern of currently unsatisfied neighboring checks. Let
\begin{equation}
    \mathcal{N}_X(i) \triangleq \{a: H_X(a,i)=1\}, \qquad
    \mathcal{N}_Z(i) \triangleq \{b: H_Z(b,i)=1\}
\end{equation}
denote the neighboring checks of VN \(i\) in the two Tanner graphs. Using fixed deterministic edge orderings, define the local residuals
\begin{equation}
    \sigma_i^X = \sum_{a\in\mathcal{N}_X(i)} \delta_a^X \, \beta_X(a,i),
    \qquad
    \sigma_i^Z = \sum_{b\in\mathcal{N}_Z(i)} \delta_b^Z \, \beta_Z(b,i),
    \label{eq:local_bitmasks}
\end{equation}
where \(\beta_X(a,i)\) and \(\beta_Z(b,i)\) are power-of-two edge weights determined by the local edge ordering. Let
\begin{equation}
    A_{\max} \triangleq \max_i \{|\mathcal{N}_X(i)|,|\mathcal{N}_Z(i)|\}.
\end{equation}
The combined local state used by the learned scheduler is
\begin{equation}
    \sigma_i \triangleq \sigma_i^X + 2^{A_{\max}}\sigma_i^Z.
    \label{eq:combined_state}
\end{equation}
At each sequential step, the decoder selects a VN according to
\begin{equation}
    i^\star = \arg\max_{i\in\mathcal{R}} \mathcal{Q}(\sigma_i,i),
    \label{eq:rl_action}
\end{equation}
where \(\mathcal{R}\) is the set of VNs that have not yet been visited in the current sweep. The selected VN is updated using the quaternary BP rule, the hard decision is refreshed, and the affected residual checks and local states are updated. In this way, the learned schedule focuses computation on parts of the Tanner graph that are most relevant to the current residual syndrome.

\section{Proposed Learned Parallel Bit-Flipping Sequential BP Decoder}
\label{sec:proposed}

\subsection{Motivation}

Learned sequential BP follows a single decoding trajectory. This is efficient, but it can be sensitive to local Pauli decisions made along the way. Under depolarizing noise, several Pauli symbols may have comparable likelihoods at a qubit. Moreover, because QLDPC codes are degenerate, a Pauli symbol that is not locally most likely may still lead to a globally successful recovery after further message passing.

Our proposed decoder adds a controlled search mechanism to learned sequential BP. It first runs RL-S for \(T_0\) iterations. If the resulting estimate satisfies the syndrome, decoding terminates. Otherwise, the decoder constructs a set of promising quaternary bit-flipping candidates using the current hard decisions, residual checks, BP messages, and local scheduling information. Each candidate changes the Pauli decision of one qubit. The top \(K\) candidates are then tested through independent continuation branches. In each branch, the decoder starts from the same configuration obtained after the initial RL-S stage, applies one forced Pauli change, refreshes the affected residual checks and BP messages, and runs RL-S for another \(T_{\rm br}\) iterations. Thus, the bit-flipping stage is a one-shot branching mechanism that provides learned sequential BP with several carefully chosen continuation paths.

\subsection{Quaternary Bit-Flipping Candidates}

Let the current Pauli decision at VN \(i\) be
\begin{equation}
    q_i^{\rm old} = \hat{q}_i .
\end{equation}
A quaternary bit-flipping candidate is a pair
\begin{equation}
    (i,q), \qquad q\in\{I,X,Y,Z\}\setminus\{q_i^{\rm old}\}.
\end{equation}
Applying this candidate changes the Pauli decision of qubit \(i\) from \(q_i^{\rm old}\) to \(q\). Depending on the old and new Pauli symbols, the corresponding binary representation may differ in the \(X\)-component, the \(Z\)-component, or both.

Define
\begin{equation}
    \Delta_i^X(q) \triangleq e^X(q_i^{\rm old}) \oplus e^X(q),
    \qquad
    \Delta_i^Z(q) \triangleq e^Z(q_i^{\rm old}) \oplus e^Z(q).
\end{equation}
If \(\Delta_i^Z(q)=1\), then the candidate changes the syndrome contribution of qubit \(i\) on its neighboring \(H_X\)-checks. If \(\Delta_i^X(q)=1\), then it changes the syndrome contribution of qubit \(i\) on its neighboring \(H_Z\)-checks.

Let
\begin{equation}
    u_i^X \triangleq \sum_{a\in\mathcal{N}_X(i)} \delta_a^X,
    \qquad
    u_i^Z \triangleq \sum_{b\in\mathcal{N}_Z(i)} \delta_b^Z
\end{equation}
be the numbers of currently unsatisfied neighboring checks in the two Tanner graphs, and let
\begin{equation}
    d_i^X \triangleq |\mathcal{N}_X(i)|, \qquad
    d_i^Z \triangleq |\mathcal{N}_Z(i)|.
\end{equation}
The local syndrome gain of candidate \((i,q)\) is
\begin{equation}
    G(i,q)
    =
    \Delta_i^Z(q)\left(2u_i^X-d_i^X\right)
    +
    \Delta_i^X(q)\left(2u_i^Z-d_i^Z\right).
    \label{eq:bf_gain}
\end{equation}
The first term measures the change in the \(H_X\)-side residual mismatch caused by modifying the \(Z\)-component of qubit \(i\), while the second term measures the corresponding change in the \(H_Z\)-side residual mismatch caused by modifying the \(X\)-component. A positive gain means that the candidate locally decreases the residual mismatch weight.

The local syndrome gain term alone may not be sufficient because it ignores the soft reliability of the candidate Pauli decision. Let
\begin{equation}
    \Lambda_i^{\max} \triangleq \max_{a\in\{I,X,Y,Z\}} \Lambda_i(a)
\end{equation}
be the largest local Pauli log-likelihood score at VN \(i\). We define the reliability cost of candidate \((i,q)\) as
\begin{equation}
    C(i,q) \triangleq \Lambda_i^{\max}-\Lambda_i(q).
    \label{eq:pauli_cost}
\end{equation}
Thus, candidates whose Pauli score is close to the best local score have small cost, whereas candidates that are less consistent with the current soft information have larger cost.

We quantize this cost using thresholds
\begin{equation}
    0<\tau_1<\tau_2<\cdots<\tau_M
\end{equation}
and define the penalty function
\begin{equation}
\phi(i,q) \triangleq
\sum_{\ell=1}^{M}
\mathbbm{1}\!\left[C(i,q)>\tau_\ell\right].
\label{eq:reliability_penalty}
\end{equation}

Finally, we include a learned lookahead term. Let \(\mathcal{A}(i,q)\) denote the set of VNs whose local residual states would change if candidate \((i,q)\) were applied. For each affected VN \(j\), let \(\sigma_j'(i,q)\) be the local state that would result from this hypothetical change. Define
\begin{equation}
    V_{\rm after}(i,q)
    \triangleq
    \max_{j\in\mathcal{A}(i,q)} \mathcal{Q}(\sigma_j'(i,q),j).
    \label{eq:v_after}
\end{equation}
This term favors candidates that lead to local states that the learned scheduler considers promising for the following RL-S continuation.

We define the overall score as
\begin{equation}
    S(i,q)
    =
    G(i,q)
    -
    \beta\phi(i,q)
    +
    \eta V_{\rm after}(i,q),
    \label{eq:candidate_score}
\end{equation}
where \(\beta\ge 0\) controls the reliability penalty and \(\eta\ge 0\) controls the learned lookahead contribution. The decoder keeps the \(K\) candidates with the largest scores among those satisfying
\begin{equation}
    G(i,q) \ge G_{\min}.
    \label{eq:min_gain}
\end{equation}

\subsection{Branch Initialization by a Forced Pauli Change}

Let the selected candidates be
\begin{equation}
    \mathcal{C}_K = \{(i_1,q_1),(i_2,q_2),\ldots,(i_K,q_K)\}.
\end{equation}
Before evaluating the branches, the decoder stores the current decoder configuration. This configuration includes the hard decisions, component estimates, posterior soft information, BP messages, residual mismatch vectors, local scheduling states, and the residual mismatch weight.

For branch \(r\), the decoder restores this saved state and applies the forced Pauli change
\begin{equation}
    \hat{q}_{i_r} \leftarrow q_r .
\end{equation}
The residual syndrome values affected by this change are updated according to the modified binary components, and the local scheduling states of neighboring VNs are refreshed.

To make the forced change visible to the following BP continuation, the incident messages of \(i_r\) are refreshed using a large finite magnitude \(L_{\rm BF}\). The sign convention is chosen so that a positive LLR favors a zero component. Thus, on the \(H_X\)-side stream, which checks the \(Z\)-component,
\begin{equation}
    L_{\rm inj}^{X{\rm -side}}(i_r)
    =
    \begin{cases}
    +L_{\rm BF}, & e^Z(q_r)=0,\\
    -L_{\rm BF}, & e^Z(q_r)=1,
    \end{cases}
\end{equation}
and on the \(H_Z\)-side stream, which checks the \(X\)-component,
\begin{equation}
    L_{\rm inj}^{Z{\rm -side}}(i_r)
    =
    \begin{cases}
    +L_{\rm BF}, & e^X(q_r)=0,\\
    -L_{\rm BF}, & e^X(q_r)=1.
    \end{cases}
\end{equation}
The finite magnitude \(L_{\rm BF}\) (in our numerical results, we use \(L_{\rm BF}=25\)) is chosen to bias the branch toward the selected Pauli decision without introducing numerical infinities.

After this initialization, the branch runs RL-S for \(T_{\rm br}\) iterations. If a branch reaches \(w=0\), it is syndrome-convergent. If no branch converges, the decoder returns the branch with the smallest final residual mismatch weight.
Algorithm~\ref{alg:lpbf_sbp} summarizes our proposed decoder.

\begin{algorithm}[!h]
\footnotesize
\caption{Learned Parallel Bit-Flipping Sequential BP Decoder}
\label{alg:lpbf_sbp}
\begin{algorithmic}[1]
\Require CSS matrices \(H_X,H_Z\); syndrome \((s^X,s^Z)\); channel LLRs; learned Q-table \(\mathcal{Q}\); parameters \(T_0,T_{\rm br},K,\beta,\eta,G_{\min},L_{\rm BF}\)
\Ensure Pauli estimate \(\hat{\bm{q}}\), convergence flag
\State Initialize quaternary BP messages and hard decisions \(\hat{\bm{q}}\)
\State Compute residual mismatches \((\delta^X,\delta^Z)\) and \(w=\|\delta^X\|_1+\|\delta^Z\|_1\)
\If{\(w=0\)}
    \State \Return \((\hat{\bm{q}},\mathrm{true})\)
\EndIf
\State Run RL-S for \(T_0\) iterations using \(\mathcal{Q}\)
\State Update \((\delta^X,\delta^Z)\), the local scheduling indices \(\{\sigma_i\}\), and \(w\)
\If{\(w=0\)}
    \State \Return \((\hat{\bm{q}},\mathrm{true})\)
\EndIf
\State \(\mathcal{C}\gets \emptyset\)
\For{each active VN \(i\)}
    \For{each \(q\in\{I,X,Y,Z\}\setminus\{\hat{q}_i\}\)}
        \State Compute the local syndrome gain \(G(i,q)\) from \eqref{eq:bf_gain}
        \If{\(G(i,q)<G_{\min}\)}
            \State \textbf{continue}
        \EndIf
        \State Compute the reliability cost \(C(i,q)\) from \eqref{eq:pauli_cost}
        \State Compute the penalty \(\phi(i,q)\) from \eqref{eq:reliability_penalty}
        \State Compute the learned lookahead value \(V_{\rm after}(i,q)\) from \eqref{eq:v_after}
        \State \(S(i,q)\gets G(i,q)-\beta\phi(i,q)+\eta V_{\rm after}(i,q)\)
        \State Insert \((i,q,S(i,q))\) into \(\mathcal{C}\)
    \EndFor
\EndFor
\State Let \(\mathcal{C}_K\) be the \(K\) highest-scoring candidates in \(\mathcal{C}\)
\If{\(\mathcal{C}_K=\emptyset\)}
    \State \Return \((\hat{\bm{q}},\mathrm{false})\)
\EndIf
\State Store the current decoder configuration as \(\mathcal{B}\)
\For{each candidate \((i_r,q_r)\in \mathcal{C}_K\) \textbf{in parallel}}
    \State Restore the branch configuration from \(\mathcal{B}\)
    \State Set \(\hat{q}_{i_r}\leftarrow q_r\)
    \State Update the affected residual syndrome values
    \State Refresh the affected local scheduling indices \(\{\sigma_i\}\)
    \State Refresh the incident BP messages using magnitude \(L_{\rm BF}\)
    \State Run RL-S for \(T_{\rm br}\) iterations
    \State Record the final estimate \(\hat{\bm{q}}^{(r)}\) and residual weight \(w_r\)
\EndFor
\If{there exists a branch \(r\) such that \(w_r=0\)}
    \State Select one successful branch \(r^\star\)
    \State \Return \((\hat{\bm{q}}^{(r^\star)},\mathrm{true})\)
\Else
    \State \(r^\star\gets \arg\min_r w_r\)
    \State \Return \((\hat{\bm{q}}^{(r^\star)},\mathrm{false})\)
\EndIf
\end{algorithmic}
\end{algorithm}

\section{Implementation and Latency}
\label{sec:complexity}

Our proposed architecture separates the decoding process into an initial learned sequential BP stage and a parallel continuation stage. The initial stage, consisting of \(T_0\) RL-S iterations, is common to all candidates. Candidate generation scans the active VNs and evaluates up to three alternative Pauli labels per VN. Since QLDPC Tanner graphs have bounded or small degrees, the gain and learned lookahead terms are computed from local neighborhoods, giving \(O(n)\) candidate-generation cost.

After the top \(K\) candidates are selected, the corresponding branches are independent. Each branch starts from the same saved decoder state, applies one forced Pauli change, and runs \(T_{\rm br}\) RL-S iterations. 
With \(K\) parallel lanes, the branches can be executed simultaneously, so the decoding depth scales as
\begin{equation}
    O(T_0+T_{\rm br}).
\end{equation}
Thus, increasing \(K\) mainly increases parallel hardware resources and memory, rather than decoding latency.

The main additional memory cost is the saved decoder state and the branch-local working memory. A fully parallel implementation replicates the branch state across lanes, while a resource-constrained implementation can time-multiplex the same branch hardware at the cost of higher latency. Hence, the proposed decoder provides a tunable tradeoff between search diversity, hardware resources, and latency.
\section{Numerical Results}
\label{sec:results}

We evaluate the proposed decoder under the depolarizing channel. The learned Q-table is trained offline as in \cite{moradi2026learning} and reused during inference. 
In our numerical results, we set \(K=25\), \(\beta=\eta=0.3\), \(G_{\min}=1\) (meaning that a candidate must produce a positive local reduction in the residual syndrome mismatch), and \(L_{\rm BF}=25\), while \(T_0\) and \(T_{\rm br}\) are varied as indicated in the figure legends. 
The performance metric is the block error rate, defined as the probability of either a logical error or a decoder failure. A decoder failure occurs when the decoder does not converge to an error estimate that satisfies the measured syndrome.

\begin{figure}[t]
  \centering
  \includegraphics[width=.9\linewidth]{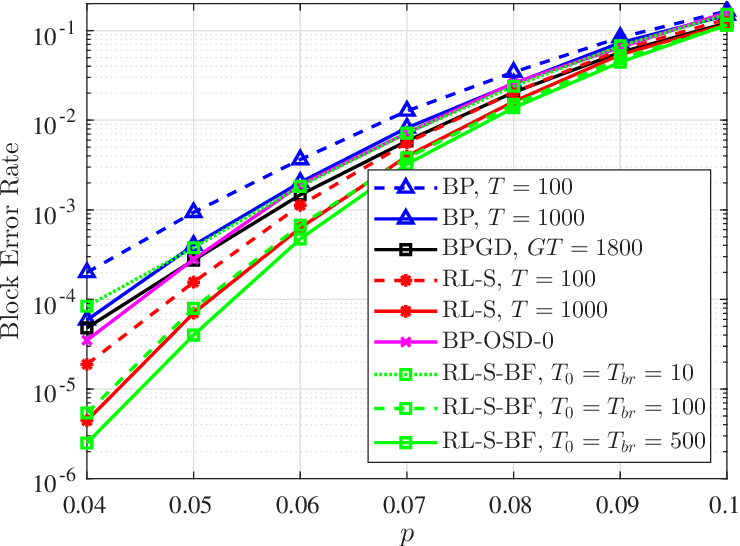}
  \caption{Block error rate of the proposed RL-S-BF decoder for the \( [[180,10,15\leq d\leq 18]] \) A5 code over the depolarizing channel. The proposed decoder is compared with BP, BPGD, RL-S, and BP-OSD-0.}
  \vspace{-.3cm}
  \label{fig:RL_S_BF_A5_FER}
\end{figure}

Fig.~\ref{fig:RL_S_BF_A5_FER} shows the block error rate of our proposed RL-S-BF decoder for the \( [[180,10,15\leq d\leq 18]] \) A5 code \cite{panteleev2021degenerate} over the depolarizing channel. The proposed decoder uses an initial RL-S stage with \(T_0=100\) sequential updates, followed by a branching stage with \(T_{\rm br}=100\) additional updates. Thus, the effective decoding depth is \(T_0+T_{\rm br}=200\). Despite this substantially smaller decoding depth, RL-S-BF achieves performance comparable to the RL-S decoder of \cite{moradi2026learning} with \(T=1000\). At \(p=0.04\), our proposed decoder improves the block error rate by approximately one order of magnitude compared with BP, BPGD \cite{yao2024bpgd}, and BP-OSD-0. In this comparison, BPGD runs BP with \(T=10\) iterations at each decimation step and decimates up to \(180\) VNs, corresponding to \(GT=1800\). These results demonstrate that our proposed branching mechanism can recover much of the error-correction performance of a much deeper RL-S decoder while requiring a significantly smaller maximum decoding depth and latency.

\begin{figure}[t]
  \centering
  \includegraphics[width=.9\linewidth]{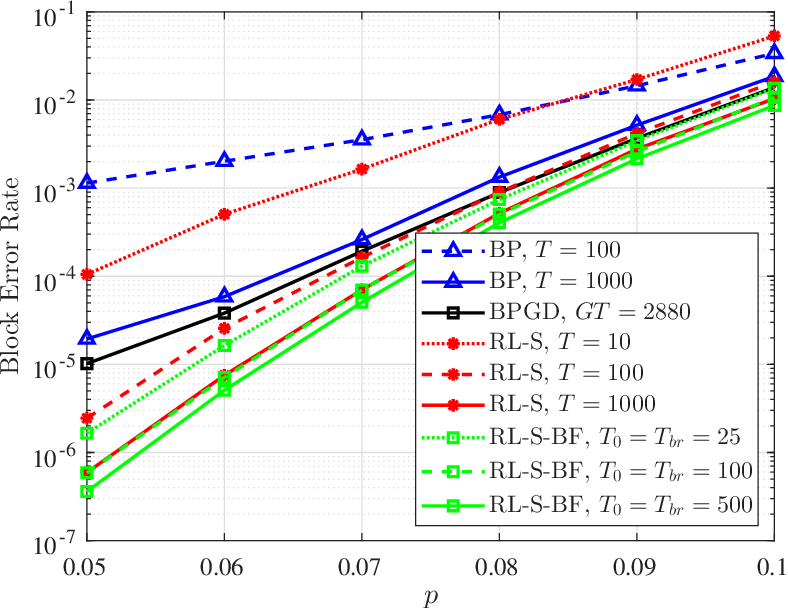}
  \caption{Block error rate of our proposed RL-S-BF decoder for the \( [[288,12,18]] \) BB code over the depolarizing channel. The proposed decoder is compared with BP, BPGD, and RL-S.}
  \vspace{-.3cm}
  \label{fig:RL_S_BF_BB288_FER}
\end{figure}

Fig.~\ref{fig:RL_S_BF_BB288_FER} plots the block error rate of our proposed RL-S-BF decoder with different maximum decoding depths for the \( [[288,12,18]] \) BB code of \cite{bravyi2024high}. With a decoding depth of \(T_0+T_{\rm br}=200\), the proposed decoder achieves performance comparable to RL-S with \(T=1000\), while providing more than one order of magnitude improvement over BP and BPGD at \(p=0.05\).

\begin{figure}[t]
  \centering
  \includegraphics[width=.9\linewidth]{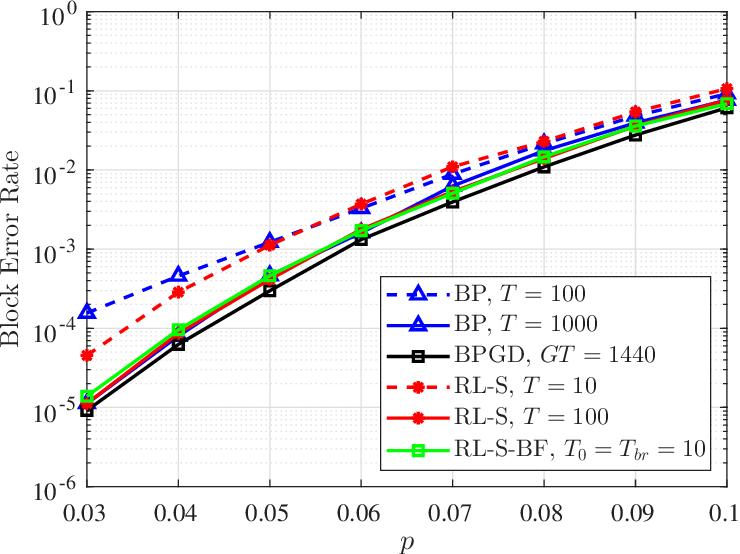}
  \caption{Block error rate of our proposed RL-S-BF decoder for the \( [[144, 12, 12 ]] \) BB code over the depolarizing channel. The proposed decoder is compared with BP, BPGD, and RL-S.}
  \vspace{-.3cm}
  \label{fig:RL_S_BF_BB144_FER}
\end{figure}

Similarly, Fig.~\ref{fig:RL_S_BF_BB144_FER} shows the block error rate of our proposed RL-S-BF decoder for the \( [[144,12,12]] \) BB code of \cite{bravyi2024high}. With a decoding depth of \(T_0+T_{\rm br}=20\), our proposed decoder achieves performance comparable to RL-S with \(T=100\), BP with \(T=1000\), and BPGD with \(GT=1140\).

As shown in \cite{moradi2026learning}, sequential BP decoding of QLDPC codes has complexity comparable to BP and substantially lower complexity than BP decoders with post-processing, such as BP-OSD. We leave a detailed complexity analysis, hardware-implementation study, and systematic optimization of the score-function parameters for future work.
\vspace{-.1cm}
\section{Conclusion}
\label{sec:conclusion}

We proposed a learned parallel bit-flipping sequential BP decoder for QLDPC codes over the depolarizing channel. The decoder first runs RL-S for a fixed number of iterations and, if the syndrome is not satisfied, constructs a set of quaternary Pauli-change candidates from the current decoder configuration. These candidates are scored using local syndrome gain, soft reliability, and learned scheduling information from the RL-S Q-table. The selected candidates are then tested through independent continuation branches, where each branch starts from the same saved configuration, applies one Pauli change, refreshes the affected residual syndrome values and BP messages, and continues RL-S decoding.
The proposed architecture adds controlled search diversity to learned sequential BP while preserving a parallel low-latency structure. Since the continuation branches are independent, several Pauli perturbations can be tested in parallel, so increasing the number of candidates mainly increases parallel resources rather than sequential decoding depth.

\vspace{-.1cm}
\section{Acknowledgment}
This work is in part supported by the National Science Foundation (NSF) under grants CNS2451268, CNS2514415, and ITE2515378, and the Office of Naval Research (ONR) under Grant N000142112472.

\bibliographystyle{IEEEtran}
\bibliography{bibliography}

\end{document}